\documentclass[12pt]{article}
\usepackage[margin=1in]{geometry}
\usepackage{amsmath, amssymb}
\usepackage{graphicx}
\usepackage{physics}
\usepackage{float}
\usepackage{tikz}
\usepackage{pgfplots}
\pgfplotsset{compat=1.18}
\usetikzlibrary{arrows.meta,positioning}
\usepackage{cite}
\usepackage{setspace}
\title{Rare Three Site Traps, Dynamic Renormalization and Slow Diffusion on Rugged Energy Landscapes}

\author{
Biman Bagchi\\
SSCU, Indian Institute of Science, Bengaluru 560012, India
}
\date{}
\begin{document}
\maketitle


\begin{abstract}
We develop a minimal theory of one-dimensional diffusion on a rugged site-energy landscape with quenched Gaussian disorder and dichotomic fluctuations. The central objects are rare three-site traps: a low-energy site flanked by two higher-energy neighbors. Because the walker cannot bypass these traps and may revisit the same environment, they strongly suppress transport. The fluctuating triplet has eight environmental configurations, from which we calculate the mean residence time and left/right escape probabilities. Temporal fluctuations weaken these bottlenecks by reducing trap persistence. In the quasi-quenched regime, repeated returns sample nearly the same trap, so the static correction to Zwanzig’s formula remains operative. Under rapid renewal, the environment changes before return, trap lifetimes shorten, and Zwanzig-like transport is recovered. The crossover reflects the interplay of hopping, renewal, and return dynamics.

\end{abstract}

\section{Introduction}

Diffusion in disordered environments is a longstanding problem in chemical physics, with applications ranging from transport in supercooled liquids, glasses, and amorphous solids to biomolecular motion on heterogeneous and conformationally fluctuating substrates [1--15]. In such systems, the diffusing particle does not move through a spatially uniform medium. Instead, it encounters a rugged energy landscape containing energetically favorable regions, unfavorable regions, and rare local configurations that can strongly delay transport. Even when the microscopic motion is Markovian, spatial heterogeneity can therefore generate broad distributions of residence times, strong deviations from simple mean-field descriptions, and pronounced dimensional dependence.

A particularly influential result was obtained by Zwanzig, who analyzed diffusion in a one-dimensional rough potential and showed that Gaussian energetic roughness produces an exponential reduction of the diffusion coefficient [16,17]. For a roughness distribution of variance $\epsilon^2$, the effective diffusion coefficient may be written as
\begin{equation}
D_{\mathrm{eff}}^{(\mathrm{Zw})}
=
D_0
\exp\!\left[-(\beta\epsilon)^2\right],
\label{eq:D_Zw}
\end{equation}
where $D_0$ is the diffusion coefficient in the absence of disorder and
$\beta=(k_{\mathrm B}T)^{-1}$. Equation~\eqref{eq:D_Zw} has become a standard reference result for diffusion in rugged landscapes. Its physical content is transparent: local energetic roughness reduces mobility by increasing the statistical weight of slow regions and activated escape events.

Zwanzig's result is most naturally interpreted as a local or effective-medium averaging of the roughness. Such an approximation can be accurate when the underlying potential varies smoothly in space, when spatial correlations suppress sharp local extrema, or when the transport process samples many microscopic fluctuations before becoming sensitive to a particular local configuration. A discrete one-dimensional lattice with independently distributed site energies, however, presents a more singular problem. In one dimension, a random walker cannot bypass an unfavorable region. It must repeatedly revisit the same local neighborhood, and therefore rare configurations with unusually long escape times can dominate the long-time diffusion coefficient. Such traps might arise in stable glasses \cite{KushalReview}

This limitation was identified explicitly by Banerjee, Biswas, Seki, and Bagchi (BBSB), who studied diffusion on a discrete one-dimensional lattice with uncorrelated Gaussian site-energy disorder \cite{Banerjee2014JCP}. They showed that Eq.~\eqref{eq:D_Zw} systematically overestimates the diffusion coefficient because it does not fully account for rare, long-lived local traps. The dominant correction arises from three-site configurations in which a low-energy central site is flanked by two higher-energy neighboring sites. Although such configurations occur with relatively small probability, their residence times can be exponentially large. Because the walker cannot circumvent them in one dimension, they act as transport bottlenecks and produce a substantial reduction of the long-time diffusion coefficient.

The resulting BBSB expression is
\begin{equation}
D_{\mathrm{BSB}}
=
D_0
\exp\!\left[-(\beta\epsilon)^2\right]
\left[
1+
\operatorname{erf}\!\left(
\frac{\beta\epsilon}{2}
\right)
\right]^{-1}.
\label{eq:D_BSB}
\end{equation}
The additional error-function factor represents the contribution of rare three-site traps that is absent from Zwanzig's local averaging. Equation~\eqref{eq:D_BSB} agrees quantitatively with continuous-time random-walk simulations of the discrete one-dimensional site-disorder model and therefore provides the appropriate quenched reference point for the present work.

The distinction between the Zwanzig and BBSB results is important for any extension to time-dependent disorder. Dynamic disorder is sometimes introduced by assigning fluctuating activation barriers directly to bonds. That construction defines a legitimate random-barrier model, but it is not equivalent to the site-energy model considered in the BBSB and related SBB studies. In the present work, the fundamental disordered variables remain the site energies. The hopping rates are derived from the instantaneous energy differences between neighboring sites, rather than being assigned independently to the connecting bonds. This distinction preserves the spatial correlations among neighboring hopping rates and retains the three-site geometry responsible for rare trapping in one dimension.

In many condensed-phase, soft-matter, and biological systems, the local energy environment is not permanently frozen. Solvation shells reorganize, hydrogen-bond networks fluctuate, local structures rearrange, and molecular conformational states interconvert. As a consequence, the energy associated with a given site may change while the particle is trapped there or while it is exploring a neighboring region. A configuration that acts as a deep trap at one instant may become less confining at a later time. Conversely, a region that is initially favorable may evolve into a slow configuration. This introduces a competition between the hopping time of the particle and the correlation time of the environment.

The central question addressed in this work is therefore:

\begin{quote}
How are the rare three-site traps of a one-dimensional rugged site-energy landscape modified when the site energies themselves fluctuate in time?
\end{quote}

The problem is formulated as a dynamically fluctuating site-disorder model. Each site energy contains a quenched random component, which represents the underlying spatial ruggedness, together with a dichotomic time-dependent component, which represents environmental reorganization. Nearest-neighbor hopping is described by Miller--Abrahams-type rates constructed from the instantaneous energy differences between adjacent sites. The disorder therefore remains a site-energy disorder, while the rates associated with the links are derived quantities. This formulation maintains instantaneous detailed balance for every frozen realization of the fluctuating landscape.

Dynamic disorder has been discussed in several related settings, including Kubo--Anderson stochastic processes, stochastic Liouville theory, dynamically disordered reaction kinetics, and hopping transport in fluctuating media [18--23]. The present problem differs from a purely local dynamic-disorder calculation because one-dimensional diffusion is controlled not only by the waiting time during a single residence event, but also by repeated returns to the same local neighborhood. In a quenched landscape, the walker encounters the same trapping configuration after every return. In a fluctuating landscape, the site energies continue to evolve during the excursion, so the trap encountered on return may be only partially correlated with the original one.

The natural local object is therefore not an isolated directed bond, but the three-site environment
$\left\{
E_{i-1}(t),E_i(t),E_{i+1}(t) \right\}$.

The total instantaneous escape rate from the central site is the sum of the left and right hopping rates,$r_i(t)
=
k_{i\rightarrow i-1}(t)
+
k_{i\rightarrow i+1}(t)$.

The residence-time statistics depend on the complete local triplet because both neighboring sites contribute to the escape process. This immediately distinguishes the present formulation from a theory in which the waiting time is assigned independently to a single directed bond.

A second issue concerns initial preparation. A particle placed randomly on a rugged landscape does not initially sample the same ensemble as a particle observed along a long stationary trajectory. In the latter case, slowly escaping sites are overrepresented because the particle spends more time in them. This distinction is physically meaningful for transient observables and for local survival probabilities. However, in a fully specified ergodic model, the asymptotic diffusion coefficient must be obtained from the stationary joint dynamics of the walker and the fluctuating landscape. The distribution of local environmental states relevant to stationary transport is therefore not an arbitrary parameter; it must be determined by the stochastic model itself.

Two related but distinct classes of models require clarification.
Stukalin and Kolomeisky obtained exact results for transport on coupled
periodic parallel lattices with stochastic interchannel transitions.
Their model provides an important benchmark for reduced two-state or
two-channel descriptions, but it does not contain quenched Gaussian
site-energy ruggedness, the Zwanzig roughness problem, or the BBSB
three-site-trap mechanism.

A recent study considered diffusion in a dynamically evolving rough
environment and emphasized that the environmental distribution sampled
by an occupied particle may differ from the bare distribution of the
landscape. Their work is physically closer to the present problem, but
their model includes reciprocal coupling between the particle and the
environment.

In the present model, by contrast, each site carries an independently
and autonomously fluctuating energy variable superposed on quenched
Gaussian disorder. The environment changes the hopping rates of the
walker, but the walker does not modify the environmental switching
dynamics. The central question is therefore how autonomous temporal
renewal modifies the lifetime and repeated recurrence of the rare
three-site traps responsible for the BBSB correction.

The objective of the present work is to construct and analyze this site-disorder problem in a form that is explicitly testable. We begin from a precisely defined fluctuating-lattice master equation for the joint probability of the walker position and the complete set of dichotomic site variables. We then develop a reduced local theory for a dynamically fluctuating three-site trap and identify the approximations involved in passing from the exact joint dynamics to a local residence-time description. The reduced theory can be compared with the exact finite-lattice master equation and with kinetic Monte Carlo simulations.

The theory is required to satisfy several limiting tests. In the limit of vanishing fluctuation rate, the landscape becomes effectively quenched and the model must approach the BBSB site-disorder result. In the limit of vanishing fluctuation amplitude, it must reduce to the ordinary static Gaussian site-energy model. In the rapid-fluctuation regime, the local dynamic component is averaged on a time scale short compared with the hopping and return times, but the resulting diffusion coefficient must be calculated from the specified model rather than assumed a priori to coincide with the Zwanzig expression.

The central physical hypothesis is that temporal fluctuations weaken one-dimensional trap dominance by reducing the persistence of rare three-site configurations. The relevant competition is not determined solely by the ratio of the local flipping rate to the bare hopping rate. It also depends on whether the environment decorrelates during the return time of the walker. Thus the dynamic renormalization of trapping is controlled jointly by the temporal correlation function of the site energies and the return-time statistics of the one-dimensional random walk.

\subsection{Three-site traps and the BBSB error-function correction}
\label{subsec:BSB_erf_origin}

The BBSB correction provides the static foundation for the dynamic theory developed below. Consider a one-dimensional lattice with independently distributed Gaussian site energies,
\begin{equation}
\rho(E)
=
\frac{1}{\sqrt{2\pi\epsilon^2}}
\exp\!\left(
-\frac{E^2}{2\epsilon^2}
\right).
\label{eq:gaussian_site_distribution}
\end{equation}
The hopping rates between nearest-neighbor sites are determined by the corresponding site-energy differences. A local three-site trap centered at site $i$ is characterized by
\begin{equation}
E_i<E_{i-1},
\qquad
E_i<E_{i+1},
\label{eq:TST_condition_intro}
\end{equation}
with particularly long residence times arising when both energy gaps,
\begin{equation}
E_{i-1}-E_i
\qquad\text{and}\qquad
E_{i+1}-E_i,
\end{equation}
are large and positive.

The mean residence time at the central site is controlled by the total escape rate
\begin{equation}
\tau_i
=
\frac{1}{
k_{i\rightarrow i-1}
+
k_{i\rightarrow i+1}
}.
\label{eq:static_residence_intro}
\end{equation}
A deep central site with only one unfavorable exit is therefore not equivalent to a deep site with two unfavorable exits. The trapping time is a property of the full three-site geometry rather than of the central site alone. This is the minimal local structure that is lost when one replaces the discrete site-energy landscape by a purely local average.

BBSB showed that the statistical contribution of these rare configurations produces the multiplicative correction in Eq.~\eqref{eq:D_BSB}. The error-function factor may therefore be interpreted as an additional one-dimensional transport resistance generated by rare three-site traps. It is absent in Zwanzig's expression because the latter averages over the roughness without retaining the detailed arrangement of neighboring site energies.

The importance of the BBSB correction extends beyond the static problem. Any theory of dynamic disorder built on the same discrete site-energy landscape must explain how temporal fluctuations modify both the probability and the lifetime of these three-site configurations. A dynamic theory that merely averages instantaneous hopping rates, without retaining the correlated left and right escape channels, cannot reproduce the mechanism responsible for Eq.~\eqref{eq:D_BSB}.

In the present work, Eq.~\eqref{eq:D_BSB} is therefore not used as an empirical fitting formula or as one endpoint of an assumed interpolation. It is used as a limiting result that the fluctuating site-energy model must recover when the environmental time scale becomes infinitely long. The manner in which this static trap correction is weakened at finite fluctuation rate will be derived and tested from the explicitly defined stochastic model developed in the following sections.

Exact results are also available for discrete stochastic models with two
coupled kinetic pathways. Stukalin and Kolomeisky derived analytical
expressions for the velocity and dispersion of particles moving on
periodic parallel lattices with interchannel coupling
\cite{StukalinKolomeisky2006}. Their model supplies a useful benchmark for
reduced two-state or two-channel descriptions of dynamic transport. The
present problem is more spatially resolved: each lattice site carries its
own fluctuating energy variable, and the two escape rates from an occupied
site share the same central-site energy. The complete stochastic state is
therefore \(P(n,\{\sigma_i\},t)\), rather than a two-component probability
associated with two globally coupled pathways.

Recent work by Makarov and Sollich has provided a particularly relevant
comparison between diffusion on static and dynamically fluctuating rough
energy landscapes \cite{MakarovSollich2026}. Their model includes the
reciprocal influence of the diffusing particle on the fluctuating
environment and shows that, under specified conditions, static and
dynamic landscapes can produce identical observable diffusivities. This
result emphasizes that the environmental distribution sampled by an
occupied particle need not coincide with the bare distribution of the
landscape. The distinction between bare and particle-conditioned sampling
is discussed further in Sec.~\ref{subsec:MS_occupied_ensembles}.



\section{Static site disorder, initial preparation, and rare traps}
\label{sec:static_initial_traps}

Before introducing temporal fluctuations of the energy landscape, it is
necessary to distinguish several statistical averages that arise even in
the static problem.

Throughout this section, the disordered variables are the \emph{site
energies}$ \{E_i\},$
rather than independently assigned bond barriers. The nearest-neighbor
hopping rates are derived from the energies of the departure and arrival
sites.

For definiteness, one may use Miller--Abrahams-type nearest-neighbor rates,
\begin{equation}
k_{i\rightarrow j}
=
k_0
\exp\left[
-\beta\max\left(0,E_j-E_i\right)
\right],
\qquad
j=i\pm1,
\label{eq:MA_static_section2}
\end{equation}
which satisfy
\begin{equation}
\frac{k_{i\rightarrow j}}{k_{j\rightarrow i}}
=
\exp\left[-\beta(E_j-E_i)\right].
\label{eq:local_DB_section2}
\end{equation}

The total escape rate from site \(i\) is therefore
\begin{equation}
r_i
=
k_{i\rightarrow i-1}
+
k_{i\rightarrow i+1},
\label{eq:total_escape_static}
\end{equation}
and the corresponding mean residence time, for a static local field is just the inverse.

The residence time is consequently a property of the three-site
configuration $(E_{i-1},E_i,E_{i+1})$.

\subsection{Random initial preparation and stationary occupation}
\label{subsec:initial_selection}

Consider a finite lattice containing \(N\) sites. A random or bare initial
preparation places the particle on a site without regard to the local
energy,
\begin{equation}
P_{\mathrm{bare}}(i)
=
\frac{1}{N}.
\label{eq:bare_initial}
\end{equation}

By contrast, a particle observed after the joint hopping process has
reached stationarity samples sites according to the stationary
occupation probability. For rates satisfying detailed balance with
respect to the site energies, the equilibrium distribution is
\begin{equation}
P_{\mathrm{eq}}(i)
=
\frac{\exp(-\beta E_i)}
{\displaystyle\sum_j\exp(-\beta E_j)}.
\label{eq:equilibrium_site_occupation}
\end{equation}

The same result may be interpreted dynamically as a residence-time bias.
If sites are encountered with a bare frequency \(P_{\mathrm{bare}}(i)\)
and the mean residence time at site \(i\) is \(\tau_i\), then the fraction
of observation time spent at that site is proportional to
$P_{\mathrm{bare}}(i)\tau_i$.
Thus one may write formally
\begin{equation}
P_{\mathrm{occ}}(i)
=
\frac{
P_{\mathrm{bare}}(i)\tau_i
}{
\displaystyle
\sum_jP_{\mathrm{bare}}(j)\tau_j
}.
\label{eq:Pocc_general}
\end{equation}
For the present site-disorder problem, however, \(\tau_i\) generally depends on the full
triplet \((E_{i-1},E_i,E_{i+1})\).

For a finite ergodic system, however, this distinction does not imply
different asymptotic diffusion constants. Once the joint stochastic
process has relaxed to its unique stationary state, the long-time
diffusion coefficient is independent of the initial preparation.

\subsection{Zwanzig's local averaging}
\label{subsec:Zwanzig_local_averaging}

Zwanzig's theory provides the classic reference point for diffusion in a
rough potential. For Gaussian energetic roughness of variance
\(\epsilon^2\), 
The detailed spatial ordering of the energy values is eliminated, and the
renormalized mobility depends only on the variance of the roughness.

This omission is especially consequential on a discrete one-dimensional
lattice. Because the walker cannot bypass a locally unfavorable region,
transport can be dominated by rare spatial configurations rather than by
the typical local roughness.

\subsection{Newman--Stein and the special role of one dimension}
\label{subsec:Newman_Stein}

The singular character of diffusion in a one-dimensional disordered
landscape was emphasized by Newman and Stein. In one dimension, a random
walker cannot move around a slow region. Any sufficiently unfavorable
local configuration encountered along the path must eventually be
crossed, often after many repeated visits.

Recurrence further amplifies this effect. After leaving a local
neighborhood, the walker has a substantial probability of returning to it.
A rare trapping configuration may therefore influence not only one
residence event but an extended sequence of excursions and returns.

\subsection{BBSB correction: three-site traps in a site-energy landscape}
\label{subsec:BBSB_traps}

Banerjee, Biswas, Seki, and Bagchi showed that Zwanzig's result
systematically overestimates diffusion on a discrete one-dimensional
lattice with uncorrelated Gaussian site energies. The leading discrepancy
arises from rare three-site traps.

A three-site trap centered at \(i\) is characterized by a low-energy
central site flanked by two higher-energy neighboring sites,
$E_i<E_{i-1}$, and $E_i<E_{i+1}$,
with especially long trapping times when both differences
$E_{i-1}-E_i$ and $E_{i+1}-E_i$ are large and positive.

For Miller--Abrahams kinetics, the two escape rates from such a central
site are
\begin{equation}
k_{i\rightarrow i-1}
=
k_0
\exp\left[-\beta(E_{i-1}-E_i)\right],
\end{equation}
and
\begin{equation}
k_{i\rightarrow i+1}
=
k_0
\exp\left[-\beta(E_{i+1}-E_i)\right],
\end{equation}
when both neighboring sites lie above \(E_i\). The total escape rate is
their sum,
\begin{equation}
r_i
=
k_0
\left\{
\exp\left[-\beta(E_{i-1}-E_i)\right]
+
\exp\left[-\beta(E_{i+1}-E_i)\right]
\right\},
\label{eq:TST_total_escape}
\end{equation}
and the residence time is
\begin{equation}
\tau_i
=
\frac{1}{r_i}.
\label{eq:TST_residence}
\end{equation}

The eventual direction of the jump is determined separately by the branching probabilities
\begin{equation}
p_{i\rightarrow i+1}
=
\frac{k_{i\rightarrow i+1}}
{k_{i\rightarrow i-1}+k_{i\rightarrow i+1}},
\end{equation}
and
\begin{equation}
p_{i\rightarrow i-1}
=
\frac{k_{i\rightarrow i-1}}
{k_{i\rightarrow i-1}+k_{i\rightarrow i+1}}.
\end{equation}

BBSB showed that the contribution of rare three-site configurations leads
to the corrected diffusion coefficient, given by Eq.2 earlier.

This distinction will be essential when dynamic disorder is introduced.
Temporal fluctuations must be applied to the site energies
$ E_{i-1}(t),\qquad E_i(t),\qquad E_{i+1}(t)$,
and the resulting left and right rates must be constructed from the
instantaneous energy differences.

\subsection{Makarov--Sollich and particle-conditioned sampling}
\label{subsec:MS_occupied_ensembles}

Recent work by Makarov and Sollich has emphasized an important distinction
between the distribution of environmental states in the landscape itself
and the distribution sampled by a particle occupying that landscape.

Let \(s\) denote a local environmental state with bare probability
\(p_0(s)\), and let \(\tau(s)\) be the corresponding mean residence time
in a model where the local state \(s\) determines the escape kinetics. The
occupied-state distribution is then
\begin{equation}
p_{\mathrm{occ}}(s)
=
\frac{
p_0(s)\tau(s)
}{
\displaystyle
\sum_{s'}p_0(s')\tau(s')
}.
\label{eq:MS_general_occupation}
\end{equation}
For the special departure-site trap model
\begin{equation}
k(s)
=
k_0
\exp[-\beta\varepsilon(s)],
\label{eq:departure_site_rate}
\end{equation}
with total escape rate \(2k(s)\), one has
\begin{equation}
\tau(s)
=
\frac{1}{2k_0}
\exp[\beta\varepsilon(s)].
\end{equation}
Equation~\eqref{eq:MS_general_occupation} then becomes
\begin{equation}
p_{\mathrm{occ}}(s)
=
\frac{
p_0(s)\exp[\beta\varepsilon(s)]
}{
\left\langle
\exp[\beta\varepsilon]
\right\rangle_0
}.
\label{eq:MS_weight_section2}
\end{equation}

For this particular model,
\begin{equation}
\sum_s
p_{\mathrm{occ}}(s)k(s)
=
\frac{k_0}{
\left\langle
\exp[\beta\varepsilon]
\right\rangle_0
}.
\label{eq:MS_average_rate}
\end{equation}

This cancellation is important, but it is model specific. It applies to a
one-site departure-rate model in which the same local variable controls
both the occupation weighting and the escape rate. It should not be
identified with a general solution of the discrete BBSB site-energy
problem.

In the BBSB model, the residence time at \(i\) depends on
\begin{equation}
(E_{i-1},E_i,E_{i+1}),
\end{equation}
and not on a single local trapping variable. Moreover, long-time transport
in one dimension includes repeated returns and correlations between
successive displacements.

The Makarov--Sollich result is relevant here because it demonstrates that
the environmental ensemble sampled by an occupied particle need not
coincide with the bare ensemble of the fluctuating landscape. This
observation motivates the distinction between a bare incoming
distribution and the stationary incoming distribution introduced below.

The two problems are nevertheless not identical. In the
Makarov--Sollich model, the particle and the environment are reciprocally
coupled, so that the presence of the particle can influence the
{\t environmental} dynamics. In the present model, the dichotomic variables
evolve autonomously and independently of the walker. The environment
affects the particle through the instantaneous hopping rates, but the
particle does not alter the flipping rate \(\nu\).

Particle-conditioned sampling should also be distinguished from the BBSB
three-site-trap correction. The former is an occupation or
residence-time weighting of environmental states. The latter arises from
the correlated two-sided escape geometry of a low-energy central site,
the Gaussian statistics of its two neighboring energies, and the
repeated recurrence of the walker in one dimension.

\subsection{Short-time jump statistics and long-time diffusion}
\label{subsec:LTT_STD}

Suppose that the particle is sampled from a
stationary distribution \(P_{\mathrm{st}}(i)\) on a fixed realization of
the site-energy landscape. Over a sufficiently short interval \(dt\), the
probability of leaving site \(i\) is
\begin{equation}
r_i\,dt
=
\left(
k_{i\rightarrow i-1}
+
k_{i\rightarrow i+1}
\right)dt.
\end{equation}
Every nearest-neighbor jump contributes \(a^2\) to the squared
displacement. Therefore
\begin{equation}
\left\langle
\Delta x^2(dt)
\right\rangle
=
a^2
\sum_i
P_{\mathrm{st}}(i)
\left(
k_{i\rightarrow i-1}
+
k_{i\rightarrow i+1}
\right)dt.
\label{eq:short_MSD_site}
\end{equation}
The corresponding instantaneous diffusion coefficient is
\begin{equation}
D(0)
=
\frac{a^2}{2}
\sum_i
P_{\mathrm{st}}(i)
\left(
k_{i\rightarrow i-1}
+
k_{i\rightarrow i+1}
\right).
\label{eq:D0_site_general}
\end{equation}

Equation~\eqref{eq:D0_site_general} is an exact short-time statement. It
does not by itself determine the long-time diffusion coefficient.

To see the difference, write the displacement after \(N(t)\) jumps as
\begin{equation}
x(t)-x(0)
=
a
\sum_{m=1}^{N(t)}
\eta_m,
\label{eq:x_jump_sum}
\end{equation}
where
\begin{equation}
\eta_m=\pm1
\end{equation}
is the direction of the \(m\)-th jump. The mean-square displacement is
then
\begin{equation}
\left\langle
[x(t)-x(0)]^2
\right\rangle
=
a^2
\left\langle N(t)\right\rangle
+
2a^2
\sum_{m<n}
\left\langle
\eta_m\eta_n
\right\rangle.
\label{eq:MSD_jump_correlation}
\end{equation}

The first term is controlled by the accumulated number of jumps. At
infinitesimal times it gives
\begin{equation}
D(0)
=
\frac{a^2}{2}
\left.
\frac{d\langle N(t)\rangle}{dt}
\right|_{t=0}.
\label{eq:D_short_jump_count}
\end{equation}
The second term contains directional correlations between distinct jumps.

The long-time diffusion coefficient is
\begin{equation}
D_\infty
=
\lim_{t\rightarrow\infty}
\frac{
\left\langle
[x(t)-x(0)]^2
\right\rangle
}{
2t
},
\label{eq:D_long_definition}
\end{equation}
or, using Eq.~\eqref{eq:MSD_jump_correlation},
\begin{equation}
D_\infty
=
\lim_{t\rightarrow\infty}
\frac{a^2}{2t}
\left[
\left\langle N(t)\right\rangle
+
2
\sum_{m<n}
\left\langle
\eta_m\eta_n
\right\rangle
\right].
\label{eq:D_long_jump}
\end{equation}

When the site energies fluctuate in time, a returning walker may encounter
a local triplet that has changed since the previous visit. The central
problem is then not the fluctuation of one isolated bond, but the temporal
persistence of the site-energy configuration
\begin{equation}
\left(
E_{i-1}(t),E_i(t),E_{i+1}(t)
\right)
\end{equation}
over the distribution of return times.



\section{Dynamically Fluctuating Site Landscape:
Benchmark Limits}
\label{sec:binary_dynamic_benchmark}

Before combining quenched Gaussian ruggedness with temporal fluctuations,
we consider a simpler reference problem with no quenched contribution.
This model isolates temporal site-energy fluctuations and provides two
analytically controlled limits.

On a one-dimensional lattice of spacing \(a\), each site energy follows
an independent symmetric telegraph process,
\begin{equation}
E_i(t)
=
\Delta \sigma_i(t),
\qquad
\sigma_i(t)=\pm1,
\label{eq:binary_dynamic_energy}
\end{equation}
with flipping rate \(\nu\). The process satisfies
\(\langle\sigma_i(t)\sigma_i(0)\rangle=e^{-2\nu|t|}\), and hence
\begin{equation}
\left\langle
E_i(t)E_i(0)
\right\rangle
=
\Delta^2\exp(-2\nu |t|).
\label{eq:binary_energy_correlation}
\end{equation}
Its correlation time is \(\tau_c=(2\nu)^{-1}\).

Nearest-neighbor hopping follows the instantaneous Miller--Abrahams rule,
\begin{equation}
k_{i\rightarrow j}(t)
=
k_0
\exp\left[
-\beta
\max\left\{
0,E_j(t)-E_i(t)
\right\}
\right],
\qquad j=i\pm1,
\label{eq:MA_binary_dynamic}
\end{equation}
which satisfies
\begin{equation}
\frac{
k_{i\rightarrow j}(t)
}{
k_{j\rightarrow i}(t)
}
=
\exp\left[
-\beta\left(E_j(t)-E_i(t)\right)
\right].
\label{eq:instantaneous_DB_binary}
\end{equation}

The primitive stochastic variables are the site energies. Adjacent rates
are correlated because they share site variables. In particular, both
\(k_{i\rightarrow i-1}(t)\) and \(k_{i\rightarrow i+1}(t)\) depend on
\(E_i(t)\), and the total escape rate is
\begin{equation}
r_i(t)
=
k_{i\rightarrow i-1}(t)
+
k_{i\rightarrow i+1}(t).
\label{eq:binary_total_escape}
\end{equation}
The benchmark is controlled by \(\beta\Delta\) and \(\nu/k_0\).

\subsection{Rapid-modulation limit}
\label{subsec:binary_fast_modulation}

For a directed hop \(i\rightarrow j\), the four equiprobable energy pairs
are \((-\Delta,-\Delta)\), \((-\Delta,+\Delta)\),
\((+\Delta,-\Delta)\), and \((+\Delta,+\Delta)\), with corresponding
rates \(k_0\), \(k_0e^{-2\beta\Delta}\), \(k_0\), and \(k_0\).
Only the low-to-high transition is activated.

When \(\nu\gg k_0\), the site energies mix before a typical hop. The
effective rate in either direction is
\begin{equation}
\overline{k}
=
k_0
\left[
\frac{3}{4}
+
\frac{1}{4}e^{-2\beta\Delta}
\right].
\label{eq:fast_average_rate}
\end{equation}
Since the left and right rates are equal, the resulting homogeneous walk
has
\begin{equation}
D_{\mathrm{fast}}
=
D_0
\left[
\frac{3}{4}
+
\frac{1}{4}e^{-2\beta\Delta}
\right],
\qquad
D_0=a^2k_0.
\label{eq:D_fast_binary}
\end{equation}

Equation~\eqref{eq:D_fast_binary} is the leading rapid-modulation result.
At finite \(k_0/\nu\), environmental memory persists during residence
events and between successive jumps. In the strong-fluctuation limit,
\(D_{\mathrm{fast}}/D_0\rightarrow3/4\): even a large instantaneous
energy difference does not form a persistent trap when renewal is
infinitely rapid.

\subsection{Frozen binary site disorder}
\label{subsec:binary_quenched_limit}

At \(\nu=0\), the site energies are frozen at \(E_i=\pm\Delta\) with
equal probabilities. For a one-dimensional reversible nearest-neighbor
walk, define the unnormalized equilibrium weight
\(\pi_i=e^{-\beta E_i}\) and the link conductance
\(c_i=\pi_i k_{i\rightarrow i+1}
=k_0e^{-\beta\max(E_i,E_{i+1})}\). The exact homogenization formula is
\begin{equation}
D_{\mathrm{q}}
=
\frac{a^2}{
\left\langle\pi_i\right\rangle
\left\langle c_i^{-1}\right\rangle
}.
\label{eq:D_quenched_general_binary}
\end{equation}

For the symmetric binary distribution,
\begin{equation}
\left\langle\pi_i\right\rangle
=
\cosh(\beta\Delta),
\qquad
\left\langle c_i^{-1}\right\rangle
=
\frac{1}{k_0}
\left[
\frac{1}{4}e^{-\beta\Delta}
+
\frac{3}{4}e^{\beta\Delta}
\right].
\label{eq:binary_averages}
\end{equation}
Therefore
\begin{equation}
D_{\mathrm{q}}
=
\frac{
D_0
}{
\displaystyle
\cosh(\beta\Delta)
\left[
\frac{1}{4}e^{-\beta\Delta}
+
\frac{3}{4}e^{\beta\Delta}
\right]
}.
\label{eq:D_quenched_binary}
\end{equation}
For \(\beta\Delta\gg1\),
\begin{equation}
D_{\mathrm{q}}
\simeq
\frac{8}{3}D_0e^{-2\beta\Delta}.
\label{eq:D_quenched_binary_strong}
\end{equation}
Thus the frozen binary landscape exhibits exponentially suppressed
transport, unlike the finite rapid-modulation limit.

\subsection{Order of limits and benchmark role}
\label{subsec:binary_order_limits}

The frozen result corresponds to setting \(\nu=0\) before taking the
long-time diffusion limit. For any strictly positive \(\nu\), the
environment eventually changes, so the limits
\(\lim_{\nu\to0^+}\lim_{t\to\infty}\langle x^2(t)\rangle/(2t)\) and
\(\lim_{t\to\infty}\lim_{\nu\to0}\langle x^2(t)\rangle/(2t)\) need not
coincide.

The two exact endpoints are
\begin{equation}
D(\nu)\longrightarrow D_{\mathrm q}
\quad\text{in the quasi-quenched regime},
\qquad
D(\nu)\longrightarrow D_{\mathrm{fast}}
\quad\text{as }\nu/k_0\rightarrow\infty.
\label{eq:binary_endpoints}
\end{equation}
The finite-\(\nu\) crossover cannot be obtained by simply averaging
static diffusion coefficients. It depends on how site-energy relaxation
overlaps with residence and return times.

The binary model therefore supplies controlled quenched and annealed
benchmarks for numerical solutions of the full joint master equation.

Exact analytical results are also available for a different class of
dynamically modulated transport models, namely coupled periodic parallel
lattices with stochastic interchannel transitions. The
Stukalin--Kolomeisky solution provides a useful benchmark for reduced
two-state or two-channel stochastic descriptions.

That model should not, however, be identified with the present
site-resolved fluctuating-energy problem. In the coupled-channel model,
the stochastic state is represented by a finite number of globally
defined pathways. In the present problem, every lattice site carries its
own dichotomic variable \(\sigma_i(t)\), the two rates leaving an
occupied site share the same central-site energy, and a previously
visited neighborhood continues to evolve while the walker is absent.
The complete stochastic state is therefore
\[
P\bigl(n,\{\sigma_i\},t\bigr),
\]
rather than a two-component probability \(P_n^{\pm}(t)\).

The Stukalin--Kolomeisky result is exact for the coupled-channel model,
but it is not the exact diffusivity of the quenched-Gaussian plus
dichotomic site-energy landscape considered here.



\section{Ground-Truth Model: Diffusion on a Dynamically
Fluctuating Gaussian Site-Energy Landscape}
\label{sec:ground_truth_dynamic_site}

We now define the stochastic model against which the subsequent local and
return-memory approximations are to be judged. The fundamental disordered
variables are the site energies; nearest-neighbor hopping rates are derived
from instantaneous site-energy differences and are not independent random
bond variables.

We consider a one-dimensional lattice of spacing \(a\). The energy of site
\(i\) is
\begin{equation}
E_i(t)
=
E_i^{(0)}
+
\Delta\sigma_i(t),
\label{eq:full_dynamic_site_energy}
\end{equation}
where \(E_i^{(0)}\) is quenched and \(\sigma_i(t)=\pm1\) is a symmetric
telegraph process.

\subsection{Quenched and dynamic components}
\label{subsec:quenched_dynamic_components}

The quenched energies are independent Gaussian variables,
\begin{equation}
\rho\!\left(E^{(0)}\right)
=
\frac{1}{\sqrt{2\pi\epsilon^2}}
\exp\left[
-\frac{\left(E^{(0)}\right)^2}{2\epsilon^2}
\right],
\label{eq:quenched_gaussian_distribution}
\end{equation}
with zero mean and covariance
\(\overline{E_i^{(0)}E_j^{(0)}}=\epsilon^2\delta_{ij}\).
Each telegraph variable flips independently with rate \(\nu\), so that
\begin{equation}
\left\langle
\sigma_i(t)\sigma_j(0)
\right\rangle_{\rm env}
=
\delta_{ij}
\exp\left(-2\nu |t|\right).
\label{eq:sigma_full_correlation}
\end{equation}
Thus the dynamic energy correlation is
\(\langle\delta E_i(t)\delta E_j(0)\rangle_{\rm env}
=\Delta^2\delta_{ij}e^{-2\nu|t|}\), with
\(\delta E_i(t)=\Delta\sigma_i(t)\) and correlation time
\(\tau_c=(2\nu)^{-1}\). The model is controlled by
\(\beta\epsilon\), \(\beta\Delta\), and \(\nu/k_0\).

\subsection{Instantaneous hopping kinetics}
\label{subsec:full_MA_rates}

For a given instantaneous landscape, nearest-neighbor hopping follows the
Miller--Abrahams rule
\begin{equation}
k_{i\rightarrow j}(t)
=
k_0
\exp\left[
-\beta
\max\left\{
0,E_j(t)-E_i(t)
\right\}
\right],
\qquad
j=i\pm1.
\label{eq:full_MA_rate}
\end{equation}
For every frozen instantaneous configuration,
\begin{equation}
\frac{
k_{i\rightarrow j}(t)
}{
k_{j\rightarrow i}(t)
}
=
\exp\left[
-\beta
\left(
E_j(t)-E_i(t)
\right)
\right].
\label{eq:full_instantaneous_DB}
\end{equation}
This instantaneous detailed balance does not imply global equilibrium
detailed balance for the autonomous walker--environment process.

The two outgoing rates share the central energy \(E_i(t)\) and are
therefore correlated. Their sum,
\begin{equation}
r_i(t)
=
k_{i\rightarrow i-1}(t)
+
k_{i\rightarrow i+1}(t),
\label{eq:full_total_escape_rate}
\end{equation}
is the total escape rate. Conditioned on escape, the left and right
branching probabilities are \(k_{i\rightarrow i-1}(t)/r_i(t)\) and
\(k_{i\rightarrow i+1}(t)/r_i(t)\), respectively.

\subsection{Exact joint master equation}
\label{subsec:full_joint_master}

For fixed quenched energies \(\{E_i^{(0)}\}\), let
\(P(n,\boldsymbol{\sigma},t)\) be the joint probability that the walker is
at site \(n\) and that the dynamic environment is
\(\boldsymbol{\sigma}=(\ldots,\sigma_{-1},\sigma_0,\sigma_1,\ldots)\).
Let \(\boldsymbol{\sigma}^{(m)}\) denote the configuration obtained by
reversing only \(\sigma_m\). The exact master equation is
\begin{align}
\frac{\partial}{\partial t}
P(n,\boldsymbol{\sigma},t)
={}&
k_{n-1\rightarrow n}(\boldsymbol{\sigma})
P(n-1,\boldsymbol{\sigma},t)
\nonumber\\
&+
k_{n+1\rightarrow n}(\boldsymbol{\sigma})
P(n+1,\boldsymbol{\sigma},t)
\nonumber\\
&-
\left[
k_{n\rightarrow n-1}(\boldsymbol{\sigma})
+
k_{n\rightarrow n+1}(\boldsymbol{\sigma})
\right]
P(n,\boldsymbol{\sigma},t)
\nonumber\\
&+
\nu
\sum_m
\left[
P(n,\boldsymbol{\sigma}^{(m)},t)
-
P(n,\boldsymbol{\sigma},t)
\right].
\label{eq:full_joint_master_explicit}
\end{align}
All environmental variables continue to evolve whether or not the walker
occupies the corresponding sites. No local renewal assumption has been
made.

The same equation may be written compactly as
\begin{equation}
\frac{\partial P}{\partial t}
=
\left[
{\cal L}_{\mathrm{hop}}
+
{\cal L}_{\mathrm{env}}
\right]P.
\label{eq:compact_joint_generator}
\end{equation}

Equations~(53) and (54) define the ground-truth stochastic model studied
in this work. The process is site resolved: each variable
\(\sigma_i(t)\) evolves independently with flipping rate \(\nu\), the
walker does not modify this environmental dynamics, and the hopping rates
are derived from instantaneous differences between neighboring site
energies.

This model is therefore distinct both from globally coupled two-channel
models and from dynamically evolving landscape models with reciprocal
particle--environment feedback. Consequently, neither the exact
diffusivity of a coupled-channel model nor a particle-conditioned
one-site averaging formula constitutes the exact solution of the present
many-site process.

\subsection{Initial preparation and transport observables}
\label{subsec:full_initial_stationary}

A factorized initial preparation has the form
\(P(n,\boldsymbol{\sigma},0)
=P_{\mathrm{pos}}^{(0)}(n)P_{\mathrm{env}}^{(0)}
(\boldsymbol{\sigma})\).
For independently sampled symmetric telegraph variables on \(N\) sites,
\(P_{\mathrm{env}}^{(0)}=2^{-N}\). The walker may be localized,
\(P_{\mathrm{pos}}^{(0)}(n)=\delta_{n,n_0}\), or placed uniformly,
\(P_{\mathrm{pos}}^{(0)}(n)=1/N\).

Even from a factorized preparation, walker and environment become
correlated. For a finite irreducible system with \(\nu>0\), the stationary
joint distribution is fixed by the master equation, although transient
observables may depend on the initial preparation.

With \(x(t)=a\,n(t)\), the diffusion coefficient for a fixed quenched
realization is
\begin{equation}
D\!\left[\{E_i^{(0)}\};\Delta,\nu\right]
=
\lim_{t\rightarrow\infty}
\frac{
\left\langle
[x(t)-x(0)]^2
\right\rangle
}{
2t
},
\label{eq:fixed_disorder_D}
\end{equation}
and the disorder-averaged coefficient is
\begin{equation}
D(\epsilon,\Delta,\nu)
=
\overline{
D\!\left[\{E_i^{(0)}\};\Delta,\nu\right]
}.
\label{eq:ground_truth_D}
\end{equation}
The finite-time coefficient
\(D(t)=\frac12\,d\langle[x(t)-x(0)]^2\rangle/dt\)
can retain preparation dependence.

\subsection{Local three-site reduction and controlled limits}
\label{subsec:full_three_site_environment}

The smallest local structure retaining the BBSB mechanism is the triplet
\({\cal T}_i(t)=[E_{i-1}(t),E_i(t),E_{i+1}(t)]\). It is instantaneously
trapping when \(E_i(t)<E_{i-1}(t)\) and \(E_i(t)<E_{i+1}(t)\).

For fixed quenched triplet energies, the dynamic state is
\(\boldsymbol{\sigma}_i=(\sigma_{i-1},\sigma_i,\sigma_{i+1})\), giving
eight environmental configurations. For configuration
\(\alpha=1,\ldots,8\), define
\begin{equation}
k_{L,\alpha}
=
k_{i\rightarrow i-1}^{(\alpha)},
\qquad
k_{R,\alpha}
=
k_{i\rightarrow i+1}^{(\alpha)},
\qquad
r_\alpha
=
k_{L,\alpha}
+
k_{R,\alpha}.
\label{eq:local_left_right_rates}
\end{equation}
This triplet retains both escape channels and the evolution of all three
site energies during a residence event.

The model has several direct checks. When \(\Delta=0\), it reduces to the
quenched Gaussian site-energy landscape. When \(\epsilon=0\), it reduces
to the fluctuating binary model of
Sec.~\ref{sec:binary_dynamic_benchmark}. When \(\nu=0\), the energies are
drawn from the static Gaussian--binary mixture
\begin{equation}
E_i
=
E_i^{(0)}
+
\Delta\sigma_i,
\qquad
\sigma_i=\pm1.
\label{eq:static_gaussian_binary_mixture}
\end{equation}
In the rapid-modulation limit \(\nu/k_0\rightarrow\infty\), the limiting
diffusion coefficient must be obtained from the specified joint dynamics
and should not be assumed to equal either the Zwanzig or BBSB result.

The local three-site theory is an approximation to
Eq.~\eqref{eq:full_joint_master_explicit}. It can be tested against
kinetic Monte Carlo simulations or finite-system solutions of the full
joint generator.



\section{Exact Local Dynamics of a Fluctuating Three-Site Trap}
\label{sec:local_three_site_dynamics}

The full joint master equation introduced in
Sec.~\ref{sec:ground_truth_dynamic_site} specifies the complete
walker--environment dynamics. The purpose of the present section is more
limited and more concrete. We isolate one visit of the walker to a
three-site neighborhood and determine exactly what happens before the
walker leaves the central site.

During this residence event, two kinds of stochastic change compete.
First, any one of the three local site energies may change because its
dichotomic variable flips. Second, the walker may escape from the central
site either to the left or to the right. The calculation below keeps
track of these two processes and yields three principal quantities: the
mean residence time, the probability of eventual escape to the left, and
the probability of eventual escape to the right. It also gives the
environmental state at the instant of escape, which is needed in the
return analysis of Sec.~\ref{sec:return_memory}.

The relevant local object is not an independently fluctuating bond, but
the three-site neighborhood
\begin{equation}
{\cal T}_i(t)
=
\left[
E_{i-1}(t),E_i(t),E_{i+1}(t)
\right].
\label{eq:Sec5_triplet}
\end{equation}
The walker is initially located at the central site \(i\). Escape may
occur to \(i-1\) or to \(i+1\), and the residence time is controlled by
the competition between these two escape channels and the continuing
fluctuation of the three local energies.

For fixed quenched values
\(E_{i-1}^{(0)}\), \(E_i^{(0)}\), and \(E_{i+1}^{(0)}\), the three site
energies evolve according to
\begin{equation}
E_\ell(t)
=
E_\ell^{(0)}
+
\Delta\sigma_\ell(t),
\qquad
\ell=i-1,i,i+1,
\label{eq:Sec5_dynamic_triplet_energies}
\end{equation}
where each \(\sigma_\ell(t)=\pm1\) is an independent symmetric telegraph
process with flipping rate \(\nu\).

The local problem considered in this section is therefore:

\begin{quote}
Given a walker initially occupying the central site of a fixed quenched
three-site environment, what are the mean residence time and the
probabilities of escape to the left and to the right when all three site
energies fluctuate in time?
\end{quote}

This local first-exit problem can be solved exactly.

\subsection{The eight environmental states}
\label{subsec:Sec5_eight_states}

At any instant, the dynamic state of the local environment is specified
by the three signs
\begin{equation}
\boldsymbol{\sigma}_i
=
\left(
\sigma_{i-1},
\sigma_i,
\sigma_{i+1}
\right),
\qquad
\sigma_\ell=\pm1.
\label{eq:Sec5_sigma_triplet}
\end{equation}
Since each of the three variables has two possible values, there are
\(2^3=8\) environmental configurations. We label them as
\begin{equation}
\begin{split}
1&=(+,+,+),\qquad
2=(+,+,-),\\
3&=(+,-,+),\qquad
4=(+,-,-),\\
5&=(-,+,+),\qquad
6=(-,+,-),\\
7&=(-,-,+),\qquad
8=(-,-,-).
\end{split}
\label{eq:Sec5_eight_state_list}
\end{equation}

The index \(\alpha=1,\ldots,8\) labels one of these environmental
configurations. For a specified state \(\alpha\), the corresponding
instantaneous energies are
\[
E_\ell^{(\alpha)}
=
E_\ell^{(0)}
+
\Delta\sigma_\ell^{(\alpha)},
\qquad
\ell=i-1,i,i+1.
\]
The eight environmental states are not eight distinct trap geometries.
A state \(\alpha\) is instantaneously trapping only when
\begin{equation}
E_i^{(\alpha)}
<
E_{i-1}^{(\alpha)},
\qquad
E_i^{(\alpha)}
<
E_{i+1}^{(\alpha)}.
\label{eq:instantaneous_trap_condition_sec5}
\end{equation}
Depending on the quenched energies and on \(\Delta\), some, all, or none
of the eight states may satisfy this condition. Nevertheless, every state
must be retained because the environment may pass through nontrapping as
well as trapping configurations before the walker escapes.

The environmental dynamics connects two states when exactly one of the
three signs changes. Thus every state is connected to three other states,
corresponding to flips at sites \(i-1\), \(i\), or \(i+1\).

We denote the \(8\times8\) environmental transition matrix by
\(\mathbf W\). Its element \(W_{\alpha\gamma}\) is the transition rate
from environmental state \(\gamma\) to environmental state \(\alpha\):
\begin{equation}
W_{\alpha\gamma}
=
\begin{cases}
\nu,
&
\text{if states \(\alpha\) and \(\gamma\) differ by one sign},
\\[4pt]
-3\nu,
&
\alpha=\gamma,
\\[4pt]
0,
&
\text{otherwise}.
\end{cases}
\label{eq:Sec5_W_definition}
\end{equation}
The off-diagonal element \(\nu\) represents a single-site flip. The
diagonal element is \(-3\nu\) because three different flips can take the
system out of any given environmental state. For each initial state
\(\gamma\), the column sum vanishes,
\(\sum_{\alpha=1}^{8}W_{\alpha\gamma}=0\), expressing conservation of
environmental probability in the absence of walker escape.

Equivalently, the environmental generator may be viewed as a sum of
three single-site contributions,
\begin{equation}
\mathbf W
=
\sum_{\ell=i-1}^{i+1}
\mathbf W^{(\ell)},
\label{eq:Sec5_W_sum}
\end{equation}
where \(\mathbf W^{(\ell)}\) generates flips of the dichotomic variable
at site \(\ell\). This form emphasizes the physical content: the three
local site variables fluctuate independently.

\subsection{Left and right escape rates}
\label{subsec:Sec5_escape_rates}

For environmental state \(\alpha\), the Miller--Abrahams escape rate
from the central site to the left is
\begin{equation}
k_{L,\alpha}
=
k_0
\exp\left[
-\beta
\max\left\{
0,
E_{i-1}^{(\alpha)}-E_i^{(\alpha)}
\right\}
\right],
\label{eq:Sec5_kL}
\end{equation}
and the corresponding escape rate to the right is
\begin{equation}
k_{R,\alpha}
=
k_0
\exp\left[
-\beta
\max\left\{
0,
E_{i+1}^{(\alpha)}-E_i^{(\alpha)}
\right\}
\right].
\label{eq:Sec5_kR}
\end{equation}

The subscript \(L\) denotes escape to site \(i-1\), whereas \(R\)
denotes escape to site \(i+1\). The index \(\alpha\) reminds us that both
rates depend on the instantaneous environmental configuration. Their sum
is the total escape rate from the central site,
\begin{equation}
r_\alpha
=
k_{L,\alpha}
+
k_{R,\alpha}.
\label{eq:Sec5_total_escape}
\end{equation}
If the environment were frozen permanently in state \(\alpha\), the mean
residence time would be \(r_\alpha^{-1}\). In the present problem,
however, the environment may change before escape, so the true residence
time must include transitions among all eight states.

It is convenient to collect the state-dependent left and right escape
rates into diagonal matrices,
\begin{equation}
\mathbf K_L
=
\operatorname{diag}
\left(
k_{L,1},k_{L,2},\ldots,k_{L,8}
\right),
\qquad
\mathbf K_R
=
\operatorname{diag}
\left(
k_{R,1},k_{R,2},\ldots,k_{R,8}
\right).
\label{eq:Sec5_K_matrices}
\end{equation}
These matrices do not transfer probability between environmental states.
Instead, they measure loss of probability by escape through the left or
right channel while the environment remains in its current state.

The total escape matrix is
\begin{equation}
\mathbf R
=
\mathbf K_L+\mathbf K_R
=
\operatorname{diag}
\left(
r_1,r_2,\ldots,r_8
\right).
\label{eq:Sec5_R}
\end{equation}
Thus \(\mathbf W\) redistributes probability among environmental states,
whereas \(\mathbf R\) removes probability from the residence event.

\subsection{Incoming distribution and local survival probabilities}
\label{subsec:Sec5_survival}

At the beginning of a residence event, the environment may be in any one
of the eight states. We denote by \(\pi_{{\rm in},\alpha}\) the
probability that the residence event begins in environmental state
\(\alpha\). These probabilities satisfy
\begin{equation}
\sum_{\alpha=1}^{8}
\pi_{{\rm in},\alpha}
=
1.
\label{eq:Sec5_initial_normalization}
\end{equation}
We collect them into the column vector
\(\boldsymbol{\pi}_{\rm in}\). The symbol \(\pi\) is used here to
distinguish this local eight-state distribution from the full
walker--environment probability
\(P(n,\boldsymbol{\sigma},t)\) introduced in Sec.~4.

The incoming distribution is not necessarily uniform. For a freshly
prepared local experiment, one may sample the three telegraph variables
independently, giving \(\pi_{{\rm in},\alpha}=1/8\). For stationary
transport, however, the incoming distribution is generated by the
previous escape and return history of the walker and is generally not
uniform.

Let \(S_\alpha(t)\) denote the probability that, at time \(t\), the walker
has not yet escaped from the central site and the environment is in state
\(\alpha\). The eight quantities \(S_\alpha(t)\) are collected into the
column vector
\begin{equation}
\mathbf S(t)
=
\begin{pmatrix}
S_1(t)\\
S_2(t)\\
\vdots\\
S_8(t)
\end{pmatrix}.
\label{eq:Sec5_survival_vector}
\end{equation}
Unlike \(\boldsymbol{\pi}_{\rm in}\), the vector \(\mathbf S(t)\) is not
normalized to unity at later times because probability is continuously
lost when the walker escapes. At \(t=0\),
\begin{equation}
S_\alpha(0)
=
\pi_{{\rm in},\alpha},
\qquad
\alpha=1,\ldots,8.
\label{eq:Sec5_initial_distribution}
\end{equation}

The survival equation may first be written componentwise:
\begin{equation}
\frac{dS_\alpha(t)}{dt}
=
\sum_{\gamma=1}^{8}
\left(
W_{\alpha\gamma}
-
R_{\alpha\gamma}
\right)
S_\gamma(t),
\label{eq:Sec5_survival_component}
\end{equation}
where \(R_{\alpha\gamma}=r_\alpha\delta_{\alpha\gamma}\). The term
containing \(W_{\alpha\gamma}\) transfers survival probability among the
eight environmental states. The term containing
\(R_{\alpha\gamma}\) removes survival probability when the walker
escapes.

In compact matrix notation, Eq.~\eqref{eq:Sec5_survival_component}
becomes
\begin{equation}
\frac{d\mathbf S(t)}{dt}
=
\left(
\mathbf W-\mathbf R
\right)
\mathbf S(t).
\label{eq:Sec5_survival_equation}
\end{equation}
The physical meaning is simple: \(\mathbf W\) changes the local
environment, whereas \(\mathbf R\) terminates the residence event.

The formal solution is
\begin{equation}
\mathbf S(t)
=
\exp\left[
(\mathbf W-\mathbf R)t
\right]
\boldsymbol{\pi}_{\rm in}.
\label{eq:Sec5_survival_solution}
\end{equation}
The exponential matrix propagates the local environment while accounting
for the possibility of escape.

The total survival probability, irrespective of the current
environmental state, is obtained by summing all eight components:
\begin{equation}
S_{\rm tot}(t)
=
\sum_{\alpha=1}^{8}
S_\alpha(t).
\label{eq:Sec5_total_survival}
\end{equation}
Because every environmental state has a positive total escape rate,
\(S_{\rm tot}(t)\) tends to zero at long times.

\subsection{Exact mean residence time}
\label{subsec:Sec5_mean_residence}

The mean residence time is the time integral of the total survival
probability,
\begin{equation}
\left\langle\tau\right\rangle
=
\int_0^\infty
S_{\rm tot}(t)\,dt.
\label{eq:Sec5_tau_definition}
\end{equation}
Using Eq.~\eqref{eq:Sec5_survival_solution}, the exact result can be
written explicitly as
\begin{equation}
\left\langle\tau\right\rangle
=
\sum_{\alpha,\gamma=1}^{8}
\left[
\left(
\mathbf R-\mathbf W
\right)^{-1}
\right]_{\alpha\gamma}
\pi_{{\rm in},\gamma}.
\label{eq:Sec5_exact_tau}
\end{equation}

The matrix inverse
\((\mathbf R-\mathbf W)^{-1}\) is the integrated survival propagator.
Its element with indices \(\alpha\gamma\) accounts for all possible
sequences of environmental flips that begin in state \(\gamma\), visit
intermediate environmental states, and contribute to survival in state
\(\alpha\) before escape. Equation~\eqref{eq:Sec5_exact_tau} therefore
does not replace the fluctuating environment by an average rate. It
treats the environmental evolution exactly during one residence event.

The quantity \(\langle\tau\rangle^{-1}\) is a local inverse residence
time. It is not, by itself, the diffusion coefficient of the full
lattice. Long-time diffusion also depends on the direction of escape,
repeated returns, the surrounding lattice, and the average over quenched
three-site configurations.

\subsection{Exact exit probabilities}
\label{subsec:Sec5_exit_probabilities}

To determine the escape direction, we first define the instantaneous
left-going probability flux. In environmental state \(\alpha\), the
surviving probability is \(S_\alpha(t)\) and the left escape rate is
\(k_{L,\alpha}\). Therefore
\begin{equation}
J_L(t)
=
\sum_{\alpha=1}^{8}
k_{L,\alpha}S_\alpha(t).
\label{eq:Sec5_left_flux}
\end{equation}
Similarly, the right-going probability flux is
\begin{equation}
J_R(t)
=
\sum_{\alpha=1}^{8}
k_{R,\alpha}S_\alpha(t).
\label{eq:Sec5_right_flux}
\end{equation}

The total probabilities of eventual escape to the left and to the right
are obtained by integrating these fluxes over time:
\begin{equation}
P_L
=
\sum_{\alpha,\gamma=1}^{8}
k_{L,\alpha}
\left[
\left(
\mathbf R-\mathbf W
\right)^{-1}
\right]_{\alpha\gamma}
\pi_{{\rm in},\gamma},
\label{eq:Sec5_PL}
\end{equation}
and
\begin{equation}
P_R
=
\sum_{\alpha,\gamma=1}^{8}
k_{R,\alpha}
\left[
\left(
\mathbf R-\mathbf W
\right)^{-1}
\right]_{\alpha\gamma}
\pi_{{\rm in},\gamma}.
\label{eq:Sec5_PR}
\end{equation}
Because escape must eventually occur through one of the two channels,
\begin{equation}
P_L+P_R=1.
\label{eq:Sec5_exit_normalization}
\end{equation}

The mean residence time and the exit probabilities answer different
questions. The quantity \(\langle\tau\rangle\) determines how long the
walker remains at the central site. The quantities \(P_L\) and \(P_R\)
determine the direction of the eventual displacement. Together they
constitute the exact solution of the local first-exit problem.

\subsection{Environmental state at the instant of escape}
\label{subsec:Sec5_exit_state}

The return problem of Sec.~\ref{sec:return_memory} requires one additional
quantity: the environmental state at the instant when the walker leaves
the central site.

Let \(\rho_{{\rm exit},\alpha}\) denote the probability that escape occurs
while the local environment is in state \(\alpha\). The incoming
probabilities \(\pi_{{\rm in},\alpha}\) describe the beginning of the
residence event, whereas the exit probabilities
\(\rho_{{\rm exit},\alpha}\) describe its end. Explicitly,
\begin{equation}
\rho_{{\rm exit},\alpha}
=
r_\alpha
\sum_{\gamma=1}^{8}
\left[
\left(
\mathbf R-\mathbf W
\right)^{-1}
\right]_{\alpha\gamma}
\pi_{{\rm in},\gamma}.
\label{eq:Sec5_exit_state}
\end{equation}
The factor \(r_\alpha\) converts the integrated survival weight in state
\(\alpha\) into the probability of escape from that state. The exit-state
distribution is normalized:
\begin{equation}
\sum_{\alpha=1}^{8}
\rho_{{\rm exit},\alpha}
=
1.
\label{eq:Sec5_exit_state_normalization}
\end{equation}

For later use, we collect the eight components into the column vector
\(\boldsymbol{\rho}_{\rm exit}\). In compact notation,
\begin{equation}
\boldsymbol{\rho}_{\rm exit}
=
\mathbf R
\left(
\mathbf R-\mathbf W
\right)^{-1}
\boldsymbol{\pi}_{\rm in}.
\label{eq:Sec5_exit_state_vector}
\end{equation}
No new physics is contained in this compact expression; it is simply the
vector form of Eq.~\eqref{eq:Sec5_exit_state}.

\subsection{Choice of the incoming environmental distribution}
\label{subsec:Sec5_initial_ensemble}

The exact local formulas are valid for any normalized incoming
distribution \(\boldsymbol{\pi}_{\rm in}\), but the physical choice of
that distribution depends on how the residence event is prepared.

For a freshly prepared local experiment, the three telegraph variables
may be sampled independently from their symmetric stationary
distribution. In that case,
\begin{equation}
\pi_{{\rm bare},\alpha}
=
\frac{1}{8},
\qquad
\alpha=1,\ldots,8.
\label{eq:Sec5_bare_distribution}
\end{equation}
This describes a walker placed at the central site without conditioning
on its previous trajectory.

For stationary transport, the relevant incoming distribution is the
distribution of local environmental states at the beginning of a
residence event after the full walker--environment process has reached
stationarity. We denote it by
\(\boldsymbol{\pi}_{\rm st,in}\). It is generated by the preceding
escape and return dynamics and is not an independent fitting parameter.

The distinction between the bare and stationary incoming distributions
can strongly affect transient residence statistics. In a finite ergodic
model, however, it does not imply multiple freely chosen asymptotic
diffusion coefficients.

\subsection{Static and rapid-fluctuation checks}
\label{subsec:Sec5_local_limits}

The exact local solution has two transparent limiting cases.

In the static limit \(\nu\rightarrow0\), environmental switching stops
and \(\mathbf W\rightarrow\mathbf 0\). Equation~\eqref{eq:Sec5_exact_tau}
then becomes
\begin{equation}
\left\langle\tau\right\rangle
=
\sum_{\alpha=1}^{8}
\frac{
\pi_{{\rm in},\alpha}
}{
r_\alpha
}.
\label{eq:Sec5_static_tau}
\end{equation}
Each residence event remains permanently in its initial environmental
state, so the result is simply an average of the eight static residence
times \(r_\alpha^{-1}\).

The corresponding exit probabilities are
\begin{equation}
P_L
=
\sum_{\alpha=1}^{8}
\pi_{{\rm in},\alpha}
\frac{k_{L,\alpha}}{r_\alpha},
\qquad
P_R
=
\sum_{\alpha=1}^{8}
\pi_{{\rm in},\alpha}
\frac{k_{R,\alpha}}{r_\alpha}.
\label{eq:Sec5_static_branching}
\end{equation}

In the opposite limit, \(\nu\gg r_\alpha\), the environment mixes among
its eight states many times before escape. To leading order, only the
uniformly averaged rates remain:
\begin{equation}
\overline{k}_L
=
\frac{1}{8}
\sum_{\alpha=1}^{8}
k_{L,\alpha},
\qquad
\overline{k}_R
=
\frac{1}{8}
\sum_{\alpha=1}^{8}
k_{R,\alpha}.
\label{eq:Sec5_fast_averaged_rates}
\end{equation}
With
\(\overline r=\overline k_L+\overline k_R\), the leading residence time
and branching probabilities are
\begin{equation}
\left\langle\tau\right\rangle
\simeq
\frac{1}{\overline r},
\qquad
P_L
\simeq
\frac{\overline{k}_L}{\overline r},
\qquad
P_R
\simeq
\frac{\overline{k}_R}{\overline r}.
\label{eq:Sec5_fast_results}
\end{equation}
Rapid switching therefore erases the identity of the initial
environmental state before escape.

These limits provide direct checks of the exact matrix inversion in
Eq.~\eqref{eq:Sec5_exact_tau}.

\subsection{Scope of the local first-exit theory}
\label{subsec:Sec5_scope}

The eight-state theory is exact for one residence event in a specified
fluctuating three-site neighborhood. It includes all three fluctuating
site energies, both competing escape channels, and every environmental
history that can occur before the walker leaves the central site.

The principal exact outputs of the section are the mean residence time
\(\langle\tau\rangle\), the left and right escape probabilities \(P_L\)
and \(P_R\), and the environmental exit-state distribution
\(\boldsymbol{\rho}_{\rm exit}\).

The local theory does not, by itself, determine the asymptotic diffusion
coefficient of the full lattice. After leaving the central site, a
one-dimensional walker may return repeatedly to the same neighborhood.
The local energies continue to evolve during the intervening excursion,
so the trap encountered on return may retain partial memory of its
earlier state. This repeated-return problem is considered next.



\section{From the BBSB Excess Time to a Dynamically Renormalized Diffusion Coefficient}
\label{sec:return_memory}

The purpose of this section is to connect the exact local three-site
calculation of Sec.~\ref{sec:local_three_site_dynamics} to an approximate
effective diffusion coefficient.

The starting point is the static BBSB correction to Zwanzig's result.
In the BBSB theory, rare three-site traps add an excess passage time to
the Zwanzig transport time. The Gaussian-wing integrations that select a
low-energy central site together with two high-energy neighboring sites
produce the familiar factor
\(\operatorname{erf}(\beta\epsilon/2)\). Thus one may write the static
BBSB passage time as
\begin{equation}
t_{\rm BBSB}
=
t_{\rm Z}
+
\Delta t_{\rm TST}^{({\rm q})},
\label{eq:Sec6_static_time_decomposition}
\end{equation}
with
\begin{equation}
\frac{
\Delta t_{\rm TST}^{({\rm q})}
}{
t_{\rm Z}
}
=
\operatorname{erf}
\left(
\frac{\beta\epsilon}{2}
\right).
\label{eq:Sec6_static_TST_fraction}
\end{equation}
Here \(t_{\rm Z}\) is the Zwanzig background passage time and
\(\Delta t_{\rm TST}^{({\rm q})}\) is the additional passage time caused
by quenched three-site traps. Since the diffusion coefficient is
inversely proportional to the passage time, Eqs.~\eqref{eq:Sec6_static_time_decomposition}
and \eqref{eq:Sec6_static_TST_fraction} give
\begin{equation}
D_{\rm BBSB}
=
\frac{
D_{\rm Z}
}{
1+
\operatorname{erf}
\left(
\frac{\beta\epsilon}{2}
\right)
}.
\label{eq:Sec6_static_BBSB}
\end{equation}

The dynamic problem differs in one essential way. The Gaussian
probability of forming a three-site trap is still determined by the
quenched-energy distribution, but the trap no longer remains fixed
indefinitely. Its local configuration changes while the walker resides
at the central site and also while the walker is away on an excursion.
The additional BBSB time must therefore be reduced by the finite
lifetime and memory of the fluctuating three-site trap.

Section~\ref{sec:local_three_site_dynamics} provides the microscopic
ingredients for this reduction. For a specified quenched triplet, the
exact mean residence time for one visit is given by
Eq.~\eqref{eq:Sec5_exact_tau}. The eight environmental states are mixed
by \(\mathbf W\), escape occurs through \(\mathbf R\), and the incoming
environmental distribution is \(\boldsymbol{\pi}_{\rm in}\). The same
section gives the exit-state distribution in
Eq.~\eqref{eq:Sec5_exit_state_vector}. Thus the local theory determines
both how long one visit lasts and the environmental state from which the
subsequent excursion begins.

The remaining task is to determine how much of the original trap
survives until the walker returns. That return-memory problem is
developed below.

\subsection{Return-memory operator}
\label{subsec:Sec6_return_memory_matrix}

Let \(\psi_{\rm ret}(t)\) be the probability density of return times to
the central site. If the local environmental distribution at departure
is \(\boldsymbol{\rho}_{\rm exit}\), then after an excursion of duration
\(t\) it becomes
\(e^{\mathbf Wt}\boldsymbol{\rho}_{\rm exit}\). Averaging over all
possible return times defines the return-memory operator
\begin{equation}
\mathbf K_{\rm ret}
=
\int_0^\infty
dt\,
\psi_{\rm ret}(t)
e^{\mathbf Wt}.
\label{eq:Sec6_Kret}
\end{equation}
The environmental distribution at the next return is therefore
\begin{equation}
\boldsymbol{\pi}_{\rm ret}
=
\mathbf K_{\rm ret}
\boldsymbol{\rho}_{\rm exit}.
\label{eq:Sec6_pi_return}
\end{equation}

Equation~\eqref{eq:Sec6_Kret} is the point at which the surrounding
lattice enters the reduced theory. The exact local three-site dynamics
is retained, but the rest of the lattice is represented by the
return-time distribution \(\psi_{\rm ret}(t)\). This is therefore a
closure of the full lattice dynamics and must ultimately be tested
against the complete joint master equation or kinetic Monte Carlo
simulation.

\subsection{Successive visits and the stationary incoming distribution}
\label{subsec:Sec6_successive_visits}

For an incoming environmental distribution
\(\boldsymbol{\pi}^{(m)}\) at the beginning of the \(m\)-th visit, define
the one-visit exit map
\begin{equation}
\mathbf Q
=
\mathbf R
\left(
\mathbf R-\mathbf W
\right)^{-1}.
\label{eq:Sec6_Q}
\end{equation}
This is the compact form of the exact exit-state result obtained in
Sec.~\ref{sec:local_three_site_dynamics}. The environmental distribution
at the end of the \(m\)-th visit is
\begin{equation}
\boldsymbol{\rho}_{\rm exit}^{(m)}
=
\mathbf Q
\boldsymbol{\pi}^{(m)}.
\label{eq:Sec6_exit_map}
\end{equation}
After the excursion and return, the incoming distribution for the next
visit is approximated by
\begin{equation}
\boldsymbol{\pi}^{(m+1)}
=
\mathbf K_{\rm ret}
\mathbf Q
\boldsymbol{\pi}^{(m)}.
\label{eq:Sec6_visit_map}
\end{equation}

A stationary sequence of visits is obtained when the incoming
distribution no longer changes from one visit to the next:
\begin{equation}
\boldsymbol{\pi}_{\rm st,in}
=
\mathbf K_{\rm ret}
\mathbf Q
\boldsymbol{\pi}_{\rm st,in},
\label{eq:Sec6_stationary_entry}
\end{equation}
with
\(\sum_{\alpha=1}^{8}\pi_{{\rm st,in},\alpha}=1\).

Equation~\eqref{eq:Sec6_stationary_entry} closes the cycle
\[
\boldsymbol{\pi}_{\rm in}
\longrightarrow
\text{residence}
\longrightarrow
\boldsymbol{\rho}_{\rm exit}
\longrightarrow
\text{return}
\longrightarrow
\boldsymbol{\pi}_{\rm in}.
\]
The stationary incoming distribution is therefore generated by the
combined residence, escape, and return dynamics; it is not chosen
independently.

Once \(\boldsymbol{\pi}_{\rm st,in}\) is known, it can be inserted into
Eq.~\eqref{eq:Sec5_exact_tau} to give the stationary mean residence time
of the fluctuating three-site neighborhood during one visit.

\subsection{Fluctuation-dependent TST excess time}
\label{subsec:Sec6_dynamic_TST_time}

The exact local residence time from Sec.~5 must now be averaged over the
quenched three-site configurations belonging to the TST sector. Denote a
quenched triplet by
\[
\Gamma
=
\left(
E_{i-1}^{(0)},
E_i^{(0)},
E_{i+1}^{(0)}
\right),
\]
and let \(P_{\rm G}(\Gamma)\) be the product of the three Gaussian
site-energy distributions. The fluctuating TST excess time may be
defined schematically as
\begin{equation}
\Delta t_{\rm TST}(\nu)
=
\int_{\rm TST}
d\Gamma\,
P_{\rm G}(\Gamma)
\left[
\left\langle\tau(\Gamma;\nu)\right\rangle_{\rm st}
-
\tau_{\rm Z}(\Gamma)
\right].
\label{eq:Sec6_dynamic_TST_excess_time}
\end{equation}
The integration domain denoted by ``TST'' selects the same low-central
and high-neighboring Gaussian-wing configurations that generate the BBSB
error-function correction in the quenched problem.
The quantity
\(\left\langle\tau(\Gamma;\nu)\right\rangle_{\rm st}\) is obtained from
Eq.~\eqref{eq:Sec5_exact_tau} using the stationary incoming distribution
determined by Eq.~\eqref{eq:Sec6_stationary_entry}.

Equation~\eqref{eq:Sec6_dynamic_TST_excess_time} provides the microscopic
definition of the fluctuation-dependent additional passage time. We do
not evaluate this Gaussian average in closed form here. Instead, we
introduce the dimensionless lifetime-renormalization factor
\begin{equation}
\Phi(\nu)
=
\frac{
\Delta t_{\rm TST}(\nu)
}{
\Delta t_{\rm TST}^{({\rm q})}
}.
\label{eq:Sec6_Phi_definition}
\end{equation}
The quenched limit gives \(\Phi(0)=1\). As environmental renewal becomes
rapid, the lifetime and return memory of the TST are reduced, so
\(\Phi(\nu)\) decreases.

The total passage time is then approximated by
\begin{equation}
t_{\rm eff}(\nu)
=
t_{\rm Z}
+
\Phi(\nu)
\Delta t_{\rm TST}^{({\rm q})}.
\label{eq:Sec6_dynamic_time_decomposition}
\end{equation}
Using Eq.~\eqref{eq:Sec6_static_TST_fraction}, the corresponding
effective diffusion coefficient is
\begin{equation}
D_{\rm eff}(\nu)
\simeq
\frac{
D_{\rm Z}
}{
1+
\Phi(\nu)
\operatorname{erf}
\left(
\frac{\beta\epsilon}{2}
\right)
}.
\label{eq:Sec6_dynamic_BBSB_general}
\end{equation}

Equation~\eqref{eq:Sec6_dynamic_BBSB_general} expresses the central
physical idea of the present theory: dynamic disorder does not remove
the Gaussian probability of forming a three-site trap, but it shortens
the additional passage time generated by that trap.

\subsection{Triplet memory and single-mode approximation}
\label{subsec:Sec6_spectral_memory}

Because \(\mathbf W\) is the sum of three independent telegraph
generators, its relaxation rates are \(0\), \(2\nu\), \(4\nu\), and
\(6\nu\). The corresponding return-memory factors are
\begin{equation}
\chi_m(\nu)
=
\int_0^\infty
dt\,
\psi_{\rm ret}(t)e^{-2m\nu t}
=
\widetilde{\psi}_{\rm ret}(2m\nu),
\qquad
m=0,1,2,3.
\label{eq:Sec6_chi_m}
\end{equation}
Here
\begin{equation}
\widetilde{\psi}_{\rm ret}(s)
=
\int_0^\infty
dt\,
e^{-st}\psi_{\rm ret}(t)
\label{eq:Sec6_return_Laplace}
\end{equation}
is the Laplace transform of the return-time density.

A single-mode approximation retains only the slowest nonstationary mode:
\begin{equation}
\chi(\nu)
=
\int_0^\infty
dt\,
\psi_{\rm ret}(t)e^{-2\nu t}.
\label{eq:Sec6_scalar_chi}
\end{equation}
For a normalized return-time density,
\begin{equation}
\chi(0)=1,
\qquad
\lim_{\nu\rightarrow\infty}
\chi(\nu)=0.
\label{eq:Sec6_chi_limits}
\end{equation}

The quantity \(\chi(\nu)\) measures the fraction of the slowest
environmental memory mode that survives during an excursion. In the
single-mode closure we identify it with the lifetime-renormalization
factor:
\begin{equation}
\Phi(\nu)
\simeq
\chi(\nu).
\label{eq:Sec6_Phi_chi}
\end{equation}
This identification is approximate. The full factor \(\Phi(\nu)\) is
defined by the eight-state residence, escape, return, and Gaussian
averaging procedure described above, whereas \(\chi(\nu)\) retains only
the slowest environmental relaxation mode.

For the illustrative exponential return-time density
\begin{equation}
\psi_{\rm ret}(t)
=
\frac{1}{\tau_{\rm ret}}
\exp\left(
-\frac{t}{\tau_{\rm ret}}
\right),
\label{eq:Sec6_exponential_return}
\end{equation}
one obtains
\begin{equation}
\chi(\nu)
=
\frac{1}{1+2\nu\tau_{\rm ret}}.
\label{eq:Sec6_exponential_chi}
\end{equation}

\subsection{Single-mode resistance closure}
\label{subsec:Sec6_resistance_closure}

More generally, let \(D_{\rm q}\) and \(D_{\rm fast}\) denote the true
quenched and rapid-modulation limits of the specified site-energy model.
A single-mode resistance interpolation is
\begin{equation}
\frac{1}{D_{\rm app}(\nu)}
=
\frac{1}{D_{\rm fast}}
+
\chi(\nu)
\left[
\frac{1}{D_{\rm q}}
-
\frac{1}{D_{\rm fast}}
\right].
\label{eq:Sec6_resistance_interpolation}
\end{equation}
This is not an exact result of the full master equation. It is a
testable closure expressing the progressive removal of the excess
resistance associated with persistent rare traps.

For the illustrative BBSB--Zwanzig crossover, we set
\(D_{\rm q}=D_{\rm BBSB}\) and \(D_{\rm fast}=D_{\rm Z}\). Using
Eq.~\eqref{eq:Sec6_Phi_chi} in
Eq.~\eqref{eq:Sec6_dynamic_BBSB_general} gives
\begin{equation}
D_{\rm app}(\nu)
=
\frac{
D_{\rm Z}
}{
1+
\chi(\nu)
\operatorname{erf}
\left(
\frac{\beta\epsilon}{2}
\right)
}.
\label{eq:illustrative_dynamic_bbsb}
\end{equation}
This formula is the direct dynamic analogue of the BBSB correction: the
static Gaussian TST contribution remains, but its associated excess time
is shortened by the return-memory factor \(\chi(\nu)\).

\subsection{Time-dependent signatures}
\label{subsec:experimental_signatures}

Dynamic trap renewal should affect not only the asymptotic diffusion
coefficient but also intermediate-time displacement statistics. Useful
observables are the self-dynamic structure factor
\begin{equation}
F_s(k,t)
=
\left\langle
\exp\left\{
ik[x(t)-x(0)]
\right\}
\right\rangle
\label{eq:Fs_experimental}
\end{equation}
and the one-dimensional non-Gaussian parameter, with
\(\Delta x(t)=x(t)-x(0)\),
\begin{equation}
\alpha_2(t)
=
\frac{
\left\langle
\Delta x^4(t)
\right\rangle
}{
3
\left\langle
\Delta x^2(t)
\right\rangle^2
}
-1.
\label{eq:NGP_one_dimension}
\end{equation}
Persistent traps should produce a larger and longer-lived non-Gaussian
signal, while increasing \(\nu\) should reduce its magnitude and
duration. This qualitative trend can be tested directly in simulations
or single-particle trajectory data.

For Fig.~1 we use
\begin{equation}
\chi(\nu)
=
\frac{1}{1+2\nu\tau_{\rm ret}},
\qquad
\tau_{\rm ret}k_0=1,
\qquad
\beta\epsilon=1.
\label{eq:Sec6_figure_parameters}
\end{equation}
These values are illustrative and are not fitted to the ground-truth
model.



\begin{figure}[H]
\centering
\begin{tikzpicture}
\begin{axis}[
    width=0.88\linewidth,
    height=0.48\linewidth,
    xmode=log,
    xmin=1e-4, xmax=1e4,
    xlabel={$\,\nu/k_0$},
    ylabel={$D_{\mathrm{eff}}/D_0$},
    legend style={at={(0.03,0.97)},anchor=north west},
    legend cell align=left,
    ymajorgrids=true,
    xmajorgrids=true,
]

\addplot[blue, thick, dashed] coordinates {(1e-4,0.20883) (1e4,0.20883)};

\addplot[red, thick, dashdotted] coordinates {(1e-4,0.0439313) (1e4,0.0439313)};

\addplot[black, thick, mark=*, mark size=1.4pt] coordinates {
(0.0001,0.0439313)
(0.000316228,0.0441018)
(0.001,0.0446386)
(0.00316228,0.0463133)
(0.01,0.0513914)
(0.0316228,0.0655492)
(0.1,0.097273)
(0.316228,0.143215)
(1,0.180332)
(3.16228,0.198613)
(10,0.205458)
(31.6228,0.207751)
(100,0.208489)
(316.228,0.208724)
(1000,0.208799)
(3162.28,0.208822)
(10000,0.20883)
};

\end{axis}
\end{tikzpicture}

\caption{
Illustrative crossover produced by the single-mode trap-memory closure.
The normalized diffusion coefficient is plotted as a function of the
environmental flipping rate \(\nu/k_0\). The lower horizontal line is the
quasi-quenched BBSB reference value, while the upper horizontal line is
the Zwanzig local-averaging reference value. The continuous curve shows
how progressive loss of return memory shortens the additional passage
time associated with persistent three-site traps. The curve is an
illustrative closure result and is not presented as the exact
finite-\(\nu\) solution of the full fluctuating site-energy master
equation.
}
\label{fig:crossover_Zw_BSB_color}
\end{figure}

Equation~\eqref{eq:illustrative_dynamic_bbsb} is a phenomenological
single-mode closure. Its purpose is to display the expected direction
and time-scale dependence of the crossover. Quantitative predictions for
the ground-truth fluctuating site-energy model require explicit
evaluation of the Gaussian average in
Eq.~\eqref{eq:Sec6_dynamic_TST_excess_time}, together with the stationary
eight-state residence and return dynamics, or direct solution of the full
joint master equation by numerical methods.


\section{Stochastic Liouville Formulation of the
Walker--Environment Dynamics}
\label{sec:stochastic_liouville}

The stochastic Liouville equation was introduced by Kubo for dynamical
systems whose evolution operator depends on a stochastic variable
\cite{Kubo1963SLE}. Related developments in fluctuation theory,
stochastic modulation, and motional narrowing were discussed by Kubo in
his treatments of the fluctuation--dissipation theorem and stochastic
line shapes \cite{Kubo1966FDT,Kubo1969LineShape}. The application of
closely related ideas to chemical rate processes in fluctuating
environments was emphasized by Zwanzig
\cite{Zwanzig1990}.

The central idea is to enlarge the state space so that both the physical
degree of freedom and the stochastic environment are included. The
dynamics is then Markovian in this joint state space, even though the
walker dynamics obtained after eliminating the environment may contain
memory.

\subsection{Additive generator in the enlarged state space}
\label{subsec:SLE_additive_generator}

The elementary transitions of the joint process are of two distinct
types. The walker may hop while the environmental configuration remains
fixed,
\[
(n,\boldsymbol{\sigma})
\longrightarrow
(n\pm1,\boldsymbol{\sigma}),
\]
with the instantaneous Miller--Abrahams rates defined in
Sec.~\ref{subsec:full_MA_rates}. Alternatively, one environmental
variable may flip while the walker remains at the same site,
\[
(n,\boldsymbol{\sigma})
\longrightarrow
(n,\boldsymbol{\sigma}^{(m)}),
\]
where \(\boldsymbol{\sigma}^{(m)}\) is obtained by reversing only
\(\sigma_m\).

Because these are distinct elementary continuous-time Markov events, the
generator of the joint process is additive:
\begin{equation}
{\cal L}
=
{\cal L}_{\rm hop}
+
{\cal L}_{\rm env}.
\label{eq:SLE_total_generator}
\end{equation}
The stochastic Liouville equation is therefore
\begin{equation}
\frac{\partial P}{\partial t}
=
\left(
{\cal L}_{\rm hop}
+
{\cal L}_{\rm env}
\right)P.
\label{eq:SLE_compact}
\end{equation}

The explicit actions of the hopping and environmental generators are
those already displayed in the ground-truth joint master equation of
Sec.~\ref{subsec:full_joint_master}. Equation~\eqref{eq:SLE_compact}
is therefore the stochastic-Liouville representation of that same
walker--environment dynamics, rather than an additional dynamical model.

The additive form of the generator does not require
\({\cal L}_{\rm hop}\) and \({\cal L}_{\rm env}\) to commute. They
generally do not commute because the hopping rates depend on the
instantaneous environmental configuration:
\begin{equation}
\left[
{\cal L}_{\rm hop},
{\cal L}_{\rm env}
\right]
\neq 0.
\label{eq:SLE_noncommuting}
\end{equation}

\subsection{Formal propagator and projection}
\label{subsec:SLE_formal_propagator}

For a fixed realization of the quenched energies, the formal solution is
\begin{equation}
P(t)
=
e^{{\cal L}t}P(0),
\label{eq:SLE_formal_solution}
\end{equation}
where \(P(t)\) denotes the full joint probability vector.

For a finite lattice of \(N\) sites, the dynamic environment has \(2^N\)
configurations, and the joint generator acts on a space of dimension
\(N2^N\). This exponential growth motivates the reduced local
descriptions used in Secs.~\ref{sec:local_three_site_dynamics} and
\ref{sec:return_memory}.

The probability of finding the walker at site \(n\), irrespective of the
environmental configuration, is
\begin{equation}
p_n(t)
=
\sum_{\boldsymbol{\sigma}}
P(n,\boldsymbol{\sigma},t).
\label{eq:SLE_reduced_walker_probability}
\end{equation}

Summation of the full joint equation over the environmental variables
removes the explicit contribution of \({\cal L}_{\rm env}\), since the
environmental generator conserves probability. The resulting equation
for \(p_n(t)\), however, is not closed. It contains quantities of the
form
\[
\sum_{\boldsymbol{\sigma}}
k_{m\rightarrow n}(\boldsymbol{\sigma})
P(m,\boldsymbol{\sigma},t),
\]
which depend on correlations between the walker position and the
instantaneous environmental configuration.

The eight-state treatment of
Sec.~\ref{sec:local_three_site_dynamics} is a local reduction of the
full stochastic Liouville problem. The approximation lies not in the
local environmental propagation, which is treated exactly within the
triplet, but in replacing the dynamics of the remaining lattice by a
return-time distribution.

\section{Dynamic disorder, trap dominance, and the glass--biology distinction}
The review by Bouchaud and Georges ~\cite{BouchaudGeorges1990} provides a unifying framework for diffusion
in disordered systems, emphasizing the dominant role of rare events, deep traps,
and broad waiting-time distributions in glassy and amorphous materials.
In quenched rugged energy landscapes---particularly in one dimension and in
marginal cases such as two dimensions---transport is controlled not by typical
barriers but by extreme fluctuations.
These rare deep traps generate long-time tails in correlation functions and can
lead to anomalous or strongly suppressed diffusion.~\cite{BouchaudGeorges1990,
Bouchaud1992,MonthusBouchaud1996}.
This physical picture underlies the breakdown of simple mean-field descriptions
such as Zwanzig's exponential renormalization and motivates discrete-lattice
corrections, including the three-site-trap mechanism identified by Banerjee, Biswas,
Seki and Bagchi \cite{Banerjee2014JCP}.

The present work introduces a complementary physical ingredient that was largely
outside the scope of the Bouchaud--Georges analysis: dynamic disorder, characterized
by a finite correlation time of the energy landscape itself.
Allowing site or barrier energies to fluctuate in time fundamentally alters the
statistics of trapping.
Deep traps are no longer permanent features of the landscape but are continually
renewed, and their lifetimes become finite even when the underlying ruggedness
is strong.
Within the present framework, this effect appears naturally through the
renormalization of the mean waiting time for escape, which governs diffusion in
one dimension.
As the environmental fluctuation rate increases, the broad waiting-time tails
characteristic of quenched disorder are progressively truncated, leading to a
smooth crossover from trap-dominated transport to a dynamically averaged regime.

This mechanism provides a natural interpretation of the crossover obtained in
this work between the Banerjee--Biswas--Seki-Bagchi limit at small fluctuation rates and
the Zwanzig mean-field limit in the fast-modulation regime.
In the quasi-quenched limit, the landscape evolves slowly compared to hopping,
and rare multi-site traps dominate transport, consistent with the Bouchaud--Georges
picture.
When the landscape fluctuates on time scales comparable to or faster than the
hopping time, trap lifetimes are renormalized downward, and diffusion is
controlled by effective escape rates rather than by the deepest static barriers.
Importantly, this crossover follows directly from the statistics of mean waiting
times and does not rely on spectral or eigenvalue arguments.

The distinction between quenched and dynamically disordered landscapes is
particularly relevant when comparing glassy systems with biological environments.
In deeply supercooled liquids and structural glasses, the energy landscape is
effectively frozen on experimental time scales, and long-lived traps (deep minima
of inherent structures) lead to
pronounced slowing down and aging phenomena.
In contrast, many biological systems---such as proteins diffusing along DNA,
conformational diffusion in biomolecules, and transport in fluctuating cellular
environments---are intrinsically dynamic.
Solvent rearrangements, conformational changes, and binding--unbinding events
continually reshape the local landscape, preventing indefinite trapping.
Transport in such systems is therefore slowed but not arrested: barriers are
renormalized rather than frozen, and long-time tails are weakened or eliminated.

From this perspective, dynamic disorder provides a physically transparent route
for reconciling rugged energy landscapes with sustained transport.
The present one-dimensional theory offers a controlled setting in which the
competition between trap dominance and trap renewal can be analyzed explicitly.
While the restriction to one dimension is deliberate---since trap effects are
maximally expressed there---the mechanism by which temporal fluctuations suppress
extreme waiting times is expected to be generic.
Extending these ideas to higher dimensions, where bypassing of traps becomes
possible and effective-medium descriptions are often invoked, remains an
important direction for future work.

\section{Dimensionality, rare events, and the pathology of one-dimensional transport}
\label{sec:dimensionality_pathology}

As discussed by Seki et al.~\cite{SekiBagchiBagchi2016_Pathological1D} diffusion 
on rugged energy landscapes exhibits 
a pronounced and highly nontrivial dependence on spatial dimensionality.
It has long been recognized that transport in one dimension is
qualitatively different from that in higher dimensions, a point emphasized
by Stein and Newman, who characterized diffusion on one-dimensional rugged
landscapes as ``pathological''.~\cite{SteinNewman1995}.
The origin of this pathology lies in the combination of geometric recurrence
and quenched disorder: a random walker in one dimension must repeatedly
revisit the same regions, so that rare but extremely slow escape events
inevitably dominate long-time transport.~\cite{Sinai1982,Derrida1983}
In the context of quenched Gaussian rugged landscapes, this behavior manifests
itself mathematically through the harmonic-mean structure of the diffusion
constant.
Because the total traversal time over a long distance is a sum of local
residence times, the slowest barriers or traps along the path exert a
disproportionate influence.
Banerjee, Biswas,  Seki, and Bagchi demonstrated  that, on a discrete
one-dimensional lattice, such rare multi-site trapping configurations lead to
a systematic breakdown of Zwanzig's mean-field expression and require the
error-function correction associated with three-site traps.~\cite{Banerjee2014JCP}
In this sense, rare events are not a perturbative correction in one dimension
but rather control the asymptotic diffusion constant.\cite{SekiBagchiBagchi2016_Pathological1D}
The situation changes qualitatively with increasing dimensionality.
Seki, Bagchi, and Bagchi investigated diffusion on rugged landscapes in higher
dimensions using an effective-medium approximation and showed that the strong
anomalies characteristic of one dimension are progressively weakened as the
dimensionality increases.
In higher dimensions, the existence of multiple alternative paths allows the
random walker to partially bypass deep local traps, reducing the dominance of
extreme waiting times.
Two dimensions plays a marginal role in this respect: recurrence still holds,
but the influence of rare events is softened compared to one dimension, while
in three dimensions transport becomes increasingly mean-field-like.

The present work is deliberately restricted to one dimension, where the
interplay between ruggedness and dynamic disorder can be analyzed in a
controlled and transparent manner.
Within this setting, dynamic disorder introduces an additional physical time
scale associated with environmental reorganization.
As shown here, temporal fluctuations of site or barrier energies renormalize
local escape kinetics by imposing a finite effective residence time even for
configurations that would act as long-lived traps in a purely quenched
landscape.
From this perspective, dynamic disorder provides a natural mechanism for
softening the one-dimensional pathology: rare trapping events are not
eliminated, but their lifetimes are truncated by environmental renewal.

These observations connect naturally to the broader framework of anomalous
transport in disordered systems reviewed by Bouchaud and Georges, where broad
distributions of waiting times and rare-event statistics are central.
In glassy systems, where disorder is effectively quenched on experimental time
scales, such rare events can lead to pronounced non-Fickian behavior and
long-time tails.
In contrast, many biological and soft-matter environments are intrinsically
dynamically disordered, so that deep traps are continually reshaped rather than
persisting indefinitely.
The present one-dimensional theory thus provides a useful reference point for
understanding how dynamic disorder competes with quenched ruggedness and
suggests that the pathological features of one-dimensional diffusion are
progressively alleviated once environmental fluctuations are taken into
account.

%

\section{Conclusions}

We have formulated diffusion on a one-dimensional rugged site-energy
landscape in which each site contains a quenched contribution and a
dichotomic fluctuating contribution. The complete walker--environment
dynamics is defined by the joint probability
\(P(n,\{\sigma_i\},t)\), with nearest-neighbor Miller--Abrahams rates
constructed from the instantaneous differences of neighboring site
energies.

Within this formulation, we solved exactly the local first-exit problem
for a walker occupying the central site of a fluctuating three-site
environment. The three dichotomic site variables generate eight
environmental configurations. Their dynamics, together with the left and
right escape rates, determines the mean residence time, the left- and
right-exit probabilities, and the environmental distribution at escape.

We also introduced a return-memory description for repeated visits to
the same three-site neighborhood. The relevant memory factor is
\[
\chi(\nu)
=
\int_0^\infty dt\,
\psi_{\rm ret}(t)e^{-2\nu t},
\]
where \(\psi_{\rm ret}(t)\) is the return-time distribution and
\(e^{-2\nu t}\) is the slowest single-site environmental relaxation
mode. In the quasi-quenched regime, the walker repeatedly encounters a
strongly correlated local environment and the BBSB rare-trap mechanism
remains effective. As the environmental renewal rate increases, memory
of the local trap is reduced over the return time and the additional
rare-trap resistance is weakened.

The present work does not provide an exact analytical expression for the
full finite-\(\nu\) diffusion coefficient of the fluctuating Gaussian
site-energy lattice. The eight-state result is exact only for one local
residence event in a specified three-site environment. The
return-memory construction is a reduced closure in which the dynamics of
the surrounding lattice is represented through a return-time
distribution. Its validity must therefore be assessed by comparison with
finite-lattice solutions of the complete joint master equation or with
kinetic Monte Carlo simulations.

Initial preparation affects transient residence statistics and
time-dependent transport, but it does not constitute an independently
adjustable parameter for the asymptotic diffusion coefficient of a
finite ergodic system. For stationary transport, the incoming
distribution of local environmental states must be determined by the
stationary joint walker--environment dynamics or by a stated closure.

Some of the prediction could possibly be tested by studying dynamics of stable glasses where rare traps could become more frequent.\cite{KushalReview}. However, to address such problems, we need to extend the present study to higher dimensions. Such work is under progress.

The present model offers an interesting extension to quantum diffusion in fluctuating lattices
studied earlier where non-Markovian or memory effects play important role. \cite{BagchiOxtoby1983}

The main physical conclusion is that temporal fluctuations modify
one-dimensional transport by reducing the persistence of rare
three-site traps over repeated returns. The crossover is therefore
controlled not only by the environmental flipping rate relative to the
bare hopping rate, but by the overlap between environmental relaxation
and the return-time statistics of the walker.


\section*{Acknowledgment}

It is a pleasure to thank Professor A. Kolomeisky for helpful discussions. 
I thank Professor K. Seki for a useful correspondence.



\end{document}